\documentclass[12pt]{iopart}
\usepackage{graphicx}
\usepackage{iopams}
\usepackage{amsbsy}
\usepackage{float}
\usepackage{amssymb}
\usepackage[makeroom]{cancel}
\usepackage[english]{babel}
\usepackage{color}
\usepackage{bm}
\usepackage{hhline}

\begin{document}
\title[Scanning gate microscopy of TMDC nanostructures]{Scanning gate microscopy mapping of edge current and branched electron flow in a transition metal dichalcogenide nanoribbon and quantum point contact}

\author{M. Prokop$^1$, D. Gut$^2$ and M. P. Nowak$^3$}
\address{$^1$ AGH University of Science and Technology, Faculty of Metals Engineering and Industrial Computer Science, al. A. Mickiewicza 30, 30-059 Krakow, Poland}
\address{$^2$AGH University of Science and Technology, Faculty of Materials Science and Ceramics, al. A. Mickiewicza 30, 30-059 Krakow, Poland}
\address{$^3$AGH University of Science and Technology, Academic Centre for Materials and Nanotechnology, al. A. Mickiewicza 30, 30-059 Krakow, Poland}

\date{\today}

\begin{abstract}
We study scanning gate microscopy (SGM) conductance mapping of a $\mathrm{MoS}_2$ zigzag ribbon exploiting tight-binding and continuum models. We show that, even though the edge modes of a pristine nanoribbon are robust to backscattering on the potential induced by the tip, the conductance mapping reveals presence of both the edge modes and the quantized spin- and valley-current carrying modes. By inspecting the electron flow from a split gate quantum point contact (QPC) we find that the mapped current flow allows to determine the nature of the quantization in the QPC as spin-orbit coupling strength affects the number of branches in which the current exits the constriction. The radial conductance oscillation fringes found in the conductance mapping reveal the presence of two possible wavevectors for the charge carriers that correspond to spin and valley opposite modes. Finally, we show that disorder induced valley mixing leads to a beating pattern in the radial fringes. 
\end{abstract}


\maketitle

\section{Introduction}
Transition Metal Dichalcogenides (TMDCs) are materials composed of a transition metal from the group VI (M: Mo, W, etc.) and a chalcogen (X: S, Se, Te) with the formula $\mathrm{MX_2}$. Being the atomic-thick semiconductor, TMDC monolayers gained recently a lot of interest as a promising material for electronic and optoelectronic application \cite{wang_electronics_2012} where they allow for the realization of e.g. FETs \cite{radisavljevic_single-layer_2011} and ultrasensitive photodetectors \cite{lopez-sanchez_ultrasensitive_2013}. Similarly to graphene, the conduction and valence band extrema in TMDCs align in two non-equivalent $K$ and $K'$ points in the Brillouin zone. This effectively gives the charge carriers an, additional to spin, degree of freedom---valley \cite{xu_spin_2014} which can be exploited for information processing, controlled by valley mixing \cite{pawlowski_valley_2018, szechenyi_impurity-assisted_2018}. In addition, the absence of inversion symmetry in the monolayers results in the strong spin-orbit coupling (SOC) which breaks the spin degeneracy of the valence and conduction band near to the energy gap \cite{zhu_giant_2011, xiao_coupled_2012} making TMDCs promising candidates for spintronics applications \cite{klinovaja_spintronics_2013, han_perspectives_2016}. A particularly important property of TMDCs nanostructures is the formation of edge states \cite{rostami_edge_2016}, which can exhibit magnetic properties when proximitized by a ferromagnet \cite{cortes_tunable_2019} or host Majorana bound states in proximity of a superconductor \cite{chu_spin-orbit-coupled_2014}. Recently, significant progress has been made in theoretical understanding of electronic transport properties of TMDCs, both in theory \cite{khoeini_peculiar_2016, ridolfi_electronic_2017} and in the experiment---thanks to the realization of locally gated nanostructures \cite{kim_genuine_2019} and split-gate QPC devices \cite{sharma_split-gated_2017, marinov_resolving_2017}.

A very powerful technique, well established for exploration of quantum transport properties in two-dimensional electron gases, is the SGM conductance mapping. In this approach a charged atomic force microscope tip scans over the sample inducing a repulsive potential in two-dimensional electron gas (2DEG) and thus scatters the propagating electrons. This technique has been applied to a variety of nanodevices \cite{sellier_imaging_2011}, mainly realized in semiconducting heterostructures where it allowed for mapping of the electron flow. The most prominent was demonstration of branched electron flow from a QPC \cite{leroy_imaging_2005, jura_electron_2009} which can be affected by mode mixing induced by Rashba SOC \cite{nowak_signatures_2014}. Most importantly, by exploiting this type of measurement the coherent nature of the electronic transport can be visualized---the electron self-interference \cite{gorini_theory_2013, kolasinski_imaging_2014} results in an appearance of radial conductance fringe patterns \cite{leroy_imaging_2005, jura_electron_2009}.

So far, SGM of monolayers was studied mainly for graphene nanoribbons \cite{mrenca_conductance_2015} and QPCs \cite{neubeck_scanning_2012, mrenca-kolasinska_imaging_2017} where it allowed for demonstration states localized within the constriction \cite{connolly_tilted_2011,  garcia_scanning_2013}. In the presence of the magnetic field this technique allowed for visualization of quantum Hall effect \cite{connolly_unraveling_2012, rajkumar_magnetic_2014}, creation of magnetic focused electron beams \cite{rajkumar_magnetic_2014,morikawa_imaging_2015,  petrovic_scanning_2017} and snake states \cite{kolasinski_imaging_2017}. Very recently, SGM of $\mathrm{MoS_2}$ started gathering attention with the first reports of visualization of electron flow  and formation of quantum dots in micrometer-sized structures \cite{bhandari_imaging_2018}. This followed the previous research that used a local probe for visualization of edge states in a few-layer $\mathrm{MoS_2}$ FET \cite{wu_uncovering_2016} using microwave impedance spectroscopy. The latter technique was also used to visualize of current flow in a $\mathrm{MoS_2}/\mathrm{WSe_2}$ heterostructure \cite{wu_visualization_2019}.

The aim of this paper is to explain how features typical to TMDCs as the presence of edge currents, strong SOC and spin-valley splitting affects SGM mapping on an example of $\mathrm{MoS_2}$. We find that in a pristine ribbon the edge current can be mapped only when the Fermi energy is tuned close to the edge band bottom. On the other, hand when the Fermi energy is tuned to the conduction band, the SGM conductance mapping reveal fans of conductance due to the presence of quantized modes in the nanoribbon rather than directly depict the current distribution \cite{aidala_imaging_2007}. By mapping the electron flow from a QPC constriction  we find that unlike as in a 2DEG in heterostructure, the number of branches is not solely dependent on the quantized conductance value but it is rather sensitive the intrinsic SOC strength. Finally, we investigate valley-mixing effect as probed by SGM mapping of the conductance oscillations.

This Paper is organized as follows: in Section 2 we describe the model used for calculations. The conductance mapping results of a pristine ribbon and a ribbon with QPC are presented in Sections 3.1 and 3.2 respectively. We summarize the work in Section 4.

\section{Model}
\begin{figure}[ht!]
\center
\includegraphics[width = 12cm]{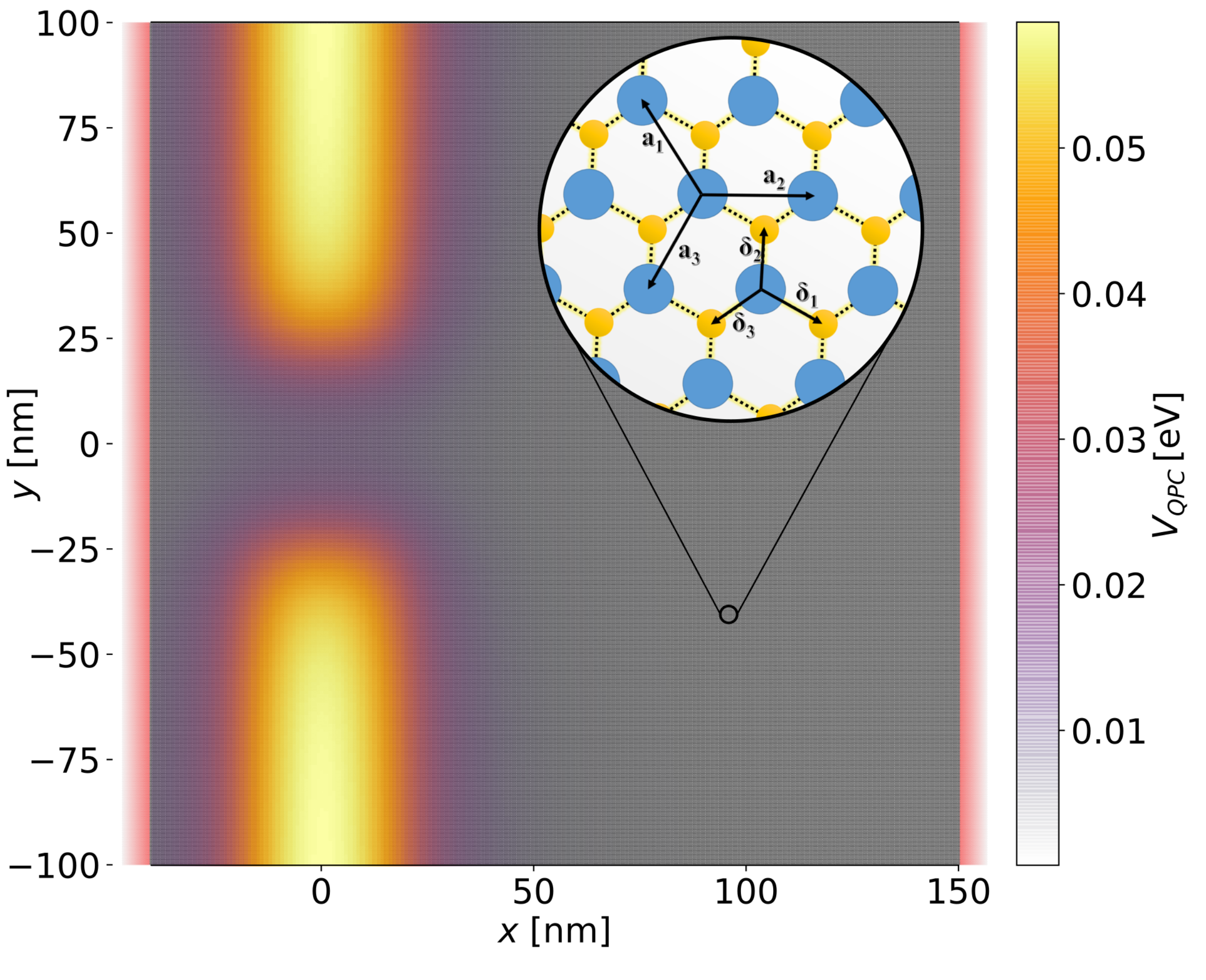}
\caption{The considered $\mathrm{MoS_2}$ ribbon. At $y=100.1$ nm and $y=-100.1$ nm the ribbon is terminated by zigzag edges formed by Mo and S atoms respectively. The gray background corresponds to the tight-binding lattice shown in the circular zoom-in. The blue and yellow circles correspond to the Mo and S atoms respectively. At the left and right edges of the sample we introduce open boundary conditions in the form of semi-infinite leads (pink colours). The colourmap depicts the QPC potential used in a part of the calculations.}
\label{system}
\end{figure}

We consider a $\mathrm{MoS_2}$ monolayer shaped into a nanoribbon as presented in Figure \ref{system}. To describe its electronic properties we adopt the tight-binding model that includes $d$ and $p$ orbitals of Mo and S atoms respectively \cite{cappelluti_tight-binding_2013}. To allow for a large scale calculations required for the description of SGM we perform reduction of the basis that casts the contribution of $p$-orbitals of the S layers into symmetric and antisymmetric combinations \cite{rostami_theory_2015, rostami_valley_2015}. The Hilbert base of the adopted model is spanned by the vector $(d_{3z^2-r^2}, d_{x^2-y^2}, d_{xy}, p_x^S, p_y^S, p_z^A)$, where the A and S indices of $p$-orbitals correspond to antisymmetric and symmetric combinations with the respect to the $z$-axis $p_i^A = 1/\sqrt2(p_i^t - p_i^b)$ $p_i^S = 1/\sqrt2(p_i^t + p_i^b)$. The index $i$ refers to spacial directions: $x,y,z$ and superscripts $t$ and $b$ indicate the top or bottom sulfur plane.

The Hamiltonian for each spin component of the considered system reads,
\begin{eqnarray}
\fl \mathbf{H} = \sum_{i,o} \left[\varepsilon^M_{i,o} a^\dagger_{i,o}a_{i,o+}\varepsilon^X_{i,o} b^\dagger_{i,o}b_{i,o}\right]
+\sum_{(i,j),o,\omega}\left[t^{MM}_{ij,o\omega}a^\dagger_{i,o}a_{j,\omega} + t^{XX}_{ij,o\omega}b^\dagger_{i,o}b_{j,\omega}\right]\nonumber \\ 
+ \sum_{(i,j),o,\omega} t^{MX}_{ij,o\omega}a^\dagger_{i,o}b_{j,\omega}  + H.c.,
\label{hamiltonian}
\end{eqnarray}
where $i,j$ iterate over lattice sites, $o,\omega$ go over atomic orbital basis and $a^\dagger$ and $b^\dagger$ are creation operators for Mo and S orbitals respectively. The first sum corresponds to the onsite energies with the elements that read,
\begin{equation}
\epsilon^M=  
\left(
\begin{array}{ccc}
\epsilon_0 + V & 0 & 0 \\
0 & \epsilon_2  +V & -i\lambda_Ms_z \\
 0 & i\lambda_Ms_z & \epsilon_2 + V\\
\end{array}\right),
\label{em}
\end{equation}
and
\begin{equation}
\epsilon^X=  \left(
\begin{array}{ccc}
\epsilon_p + t_{xx} + V & -i\frac{\lambda_X}{2}s_z & 0 \\
 i\frac{\lambda_X}{2}s_z & \epsilon_p + t_{yy} + V & 0 \\
 0 & 0 & \epsilon_z - t_{zz} + V\\
\end{array}\right).
\label{ex}
\end{equation}
In the above we include external, position dependent potential $V = V_{QPC} + V_t$ and where $s_z$ equals 1(-1) for spin up (down) component. 

The second and third sum in Hamiltonian (\ref{hamiltonian}) correspond to the hopping elements between intra- and inter-atomic orbitals respectively. The mapping of $i,j$ coordinates into the hopping elements is presented in the inset to Figure \ref{system}. The lattice spacing is 0.319 nm. The hopping matrices for the intra- and inter-lattice hoppings are,
\begin{equation}
t _1^{XX}= \frac{1} {4}\left( 
\begin{array}{ccc}
3V_{pp\pi} + V_{pp\sigma} & \sqrt3(V_{pp\pi} - V_{pp\sigma}) & 0 \\
\sqrt3(V_{pp\pi} - V_{pp\sigma}) & V_{pp\pi} + 3V_{pp\sigma} & 0 \\
0 & 0 & 4V_{pp\pi} \\
\end{array}\right),
\end{equation}
\begin{equation}
t _2^{XX}= \left(
\begin{array}{ccc}
V_{pp\sigma} & 0 & 0 \\
0 & V_{pp\pi} & 0 \\
0 & 0 & V_{pp\pi} \\
\end{array}\right),
\end{equation}
\begin{equation}
t _3^{XX}= \frac{1} {4} \left(
\begin{array}{ccc}
3V_{pp\pi} + V_{pp\sigma} & -\sqrt3(V_{pp\pi} - V_{pp\sigma}) & 0 \\
-\sqrt3(V_{pp\pi} - V_{pp\sigma}) & V_{pp\pi} + 3V_{pp\sigma} & 0 \\
0 & 0 & 4V_{pp\pi} \\
\end{array}\right),
\end{equation}

\begin{equation}
\fl t _1^{MM}= \frac{1} {4} \left(
\begin{array}{ccc}
3V_{dd\delta} + V_{dd\sigma} & \frac{\sqrt3}{2}(-V_{dd\delta} + V_{dd\sigma}) & -\frac3 2(V_{dd\delta} - V_{dd\sigma}) \\
\frac{\sqrt3}{2}(-V_{dd\delta} + V_{dd\sigma}) & \frac{1}{4}(V_{dd\delta} + 12V_{dd\pi} + 3V_{dd\sigma}) & \frac{\sqrt3} 4(V_{dd\delta} - 4V_{dd\pi} + 3V_{dd\sigma}) \\
-\frac3 2(V_{dd\delta} - V_{dd\sigma}) & \frac{\sqrt3}{4}(V_{dd\delta} - 4V_{dd\pi} + 3V_{dd\sigma}) & \frac1 4(3V_{dd\delta} + 4V_{dd\pi} + 9V_{dd\sigma}) \\
\end{array}\right),
\end{equation}
\begin{equation}
\fl t _2^{MM}= \frac{1} {4} \left(
\begin{array}{ccc}
3V_{dd\delta} + V_{dd\sigma} & \sqrt3(V_{dd\delta} - V_{dd\sigma}) & 0 \\
\sqrt3(V_{dd\delta} - V_{dd\sigma}) & V_{dd\delta} + 3V_{dd\sigma} & 0 \\
0 & 0 & 4V_{dd\pi} \\
\end{array}\right),
\end{equation}
\begin{equation}
\fl t _3^{MM}= \frac{1} {4} \left(
\begin{array}{ccc}
3V_{dd\delta} + V_{dd\sigma} & \frac{\sqrt3}{2}(-V_{dd\delta} + V_{dd\sigma}) & \frac3 2(V_{dd\delta} - V_{dd\sigma}) \\
\frac{\sqrt3}{2}(-V_{dd\delta} + V_{dd\sigma}) & \frac{1}{4}(V_{dd\delta} + 12V_{dd\pi} + 3V_{dd\sigma}) & -\frac{\sqrt3} 4(V_{dd\delta} - 4V_{dd\pi} + 3V_{dd\sigma}) \\
\frac3 2(V_{dd\delta} - V_{dd\sigma}) & -\frac{\sqrt3}{4}(V_{dd\delta} - 4V_{dd\pi} + 3V_{dd\sigma}) & \frac1 4(3V_{dd\delta} + 4V_{dd\pi} + 9V_{dd\sigma}) \\
\end{array}\right),
\end{equation}
\begin{equation}
\fl t _1^{MX}= \frac{\sqrt 2} {7\sqrt7} \left(
\begin{array}{ccc}
-9V_{pd\pi} + \sqrt3V_{pd\sigma} & 3\sqrt3V_{pd\pi} - V_{pd\sigma} & 12V_{pd\pi} + \sqrt3V_{pd\sigma}\\
5\sqrt3V_{pd\pi} + 3V_{pd\sigma} & 9V_{pd\pi} - \sqrt3V_{pd\sigma} & -2\sqrt3V_{pd\pi} + 3V_{pd\sigma}\\
-V_{pd\pi} - 3\sqrt3V_{pd\sigma} & 5\sqrt3V_{pd\pi} + 3V_{pd\sigma} & 6V_{pd\pi} - 3\sqrt3V_{pd\sigma}\\
\end{array}\right),
\end{equation}
\begin{equation}
\fl t _2^{MX}= \frac{\sqrt 2} {7\sqrt7} \left(
\begin{array}{ccc}
0 & -6\sqrt3V_{pd\pi} + 2V_{pd\sigma} & 12V_{pd\pi} + \sqrt3V_{pd\sigma}\\
0 & -6V_{pd\pi} - 4\sqrt3V_{pd\sigma} & 4\sqrt3V_{pd\pi} - 6V_{pd\sigma}\\
14V_{pd\pi} & 0 & 0\\
\end{array}\right),
\end{equation}
\begin{equation}
\fl t _3^{MX}= \frac{\sqrt 2} {7\sqrt7} \left(
\begin{array}{ccc}
9V_{pd\pi} - \sqrt3V_{pd\sigma} & 3\sqrt3V_{pd\pi} - V_{pd\sigma} & 12V_{pd\pi} + \sqrt3V_{pd\sigma}\\
-5\sqrt3V_{pd\pi} - 3V_{pd\sigma} & 9V_{pd\pi} - \sqrt3V_{pd\sigma} & -2\sqrt3V_{pd\pi} + 3V_{pd\sigma}\\
-V_{pd\pi} - 3\sqrt3V_{pd\sigma} & -5\sqrt3V_{pd\pi} - 3V_{pd\sigma} & -6V_{pd\pi} + 3\sqrt3V_{pd\sigma}\\
\end{array}\right).
\end{equation}

In (\ref{em}) and (\ref{ex}) $\lambda_M$ and $\lambda_X$ correspond to intrinsic SOC parameters which we choose after \cite{kosmider_large_2013}, with the modification of $\lambda_M = -0.086$ eV and $\lambda_S = 0.013$ eV which assures the $3$ meV spin-orbit splitting in the conduction band minimum of and the crossing of the conduction bands as found in \cite{kormanyos_spin-orbit_2014}. We adopt the following Slater-Koster parameters given in eV units: $V_{pd\sigma}$ = 3.689, $V_{pd\pi}$ = -1.241, $V_{dd\sigma}$ =-0.895, $V_{dd\pi}$ = 0.252, $V_{dd\delta}$ = 0.228, $V_{pp\sigma}$  = 1.225, $V_{pp\pi}$ = -0.467, $\epsilon_0$ =-1.094, $\epsilon_2$ = -1.512, $\epsilon_p$ = -3.560, $\epsilon_z$ = -6.886 \cite{rostami_theory_2015}.

In our work we consider an external potential induced in the monolayer that defines the QPC constriction modeled as a split-gate \cite{davies_modeling_1995}:
\begin{eqnarray}
\fl V_{\mathrm{QPC}}(x,y) = \frac{V_g}{\pi}\left[ \arctan\left(\frac{W+x}{d}\right)+\arctan\left(\frac{W-x}{d}\right)\right] \nonumber \\ 
-g(S+y,W+x)-g(S+y,W-x)\nonumber \\
-g(S-y,W+x)-g(S-y,W-x),
\label{QPC_potential}
\end{eqnarray}
with $g(u,v)=\frac{1}{2\pi}\arctan(\frac{uv}{dR})$ and $R=\sqrt{u^2+v^2+d^2}$, where $W$ and $S$ control the span of the potential in the $x$ and $y$ directions respectively and where $d$ is the parameter that controls its smoothness. We take $W=20$ nm, $S=20$ nm and $d=15$ nm. The resulting potential for $V_g=0.1$ eV is plotted in Figure \ref{system} on the colourmap.

For the SGM conductance mapping we model potential induced by the tip following \cite{szafran_scanning_2011} that well approximates the SGM potential generated in atomic-thick materials \cite{zebrowski_aharonov-bohm_2018},
\begin{equation}
V_{\mathrm{t}}(x,y) = \frac{V_{\mathrm{tip}}}{1+\frac{(x-x_t)^2+(y-y_t)^2}{\gamma^2}},
\end{equation}
with $V_{\mathrm{tip}} = 0.1$ eV and the effective width of
the tip potential $\gamma = 5$ nm for a pristine ribbon and $\gamma = 1$ nm for the system with a QPC.

We consider linear response regime at zero temperature where the conductance is obtained from Landauer formula. The scattering matrix is calculated using wave-function matching method implemented in Kwant package \cite{groth_kwant:_2014}. The conductance maps and plots were obtained using Adaptive package \cite{nijholt_textitadaptive:_2019}.

\section{Results}
\subsection{Pristine zigzag wire}
\begin{figure}[ht!]
\center
\includegraphics[width = 10cm]{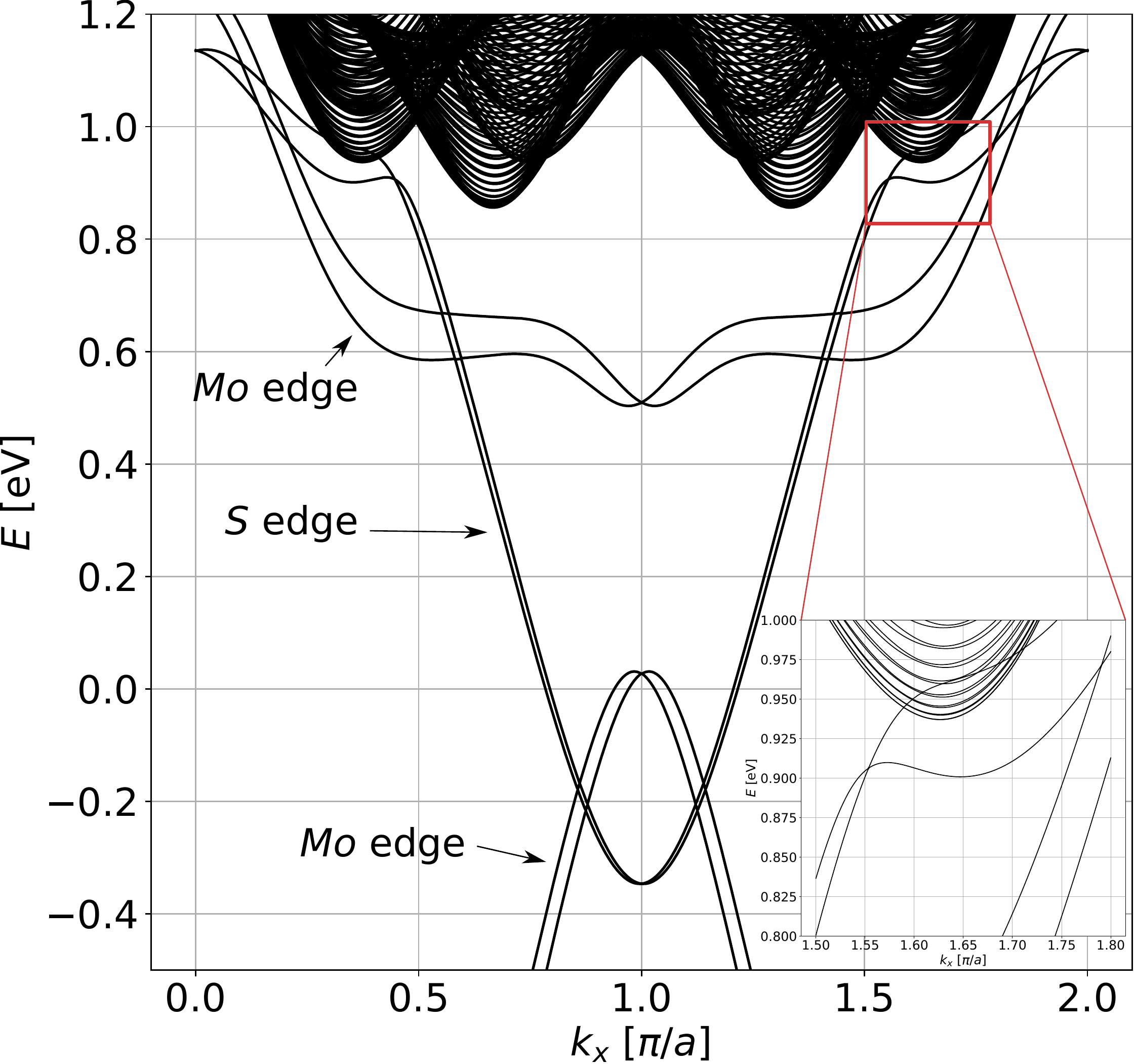}
\caption{Dispersion relation of a $\mathrm{MoS_2}$ zigzag nanoribbon with denoted bands corresponding to the edge modes. The inset shows zoom-in on the bands of $Q$ polarized modes and four bands corresponding to the edge modes.}
\label{bandstructure}
\end{figure}

In the considered system, the lattice in the $y$ direction is terminated by non-equivalent atoms. At the bottom edge of the ribbon is terminated by S atoms, while on the top it is terminated by Mo atoms. Those terminations constitute two zigzag edges of the ribbon that have different electronic structure. In Figure \ref{bandstructure} we present the band structure of a 24 nm width zigzag nanoribbon. With the arrows we denote three spin-split bands that correspond to the modes located at the Mo and S edges of the ribbon. In the top part of the figure we observe sets of bands that correspond to the current polarized in $K$, $K'$ and $Q$ valleys \cite{gut_valley_2019}. Note that changing of the width of the ribbon does not modify the structure of the edge bands until critical width of 2 nm is reached and the edge modes start to mix. 

\begin{figure}[ht!]
\center
\includegraphics[width = 10cm]{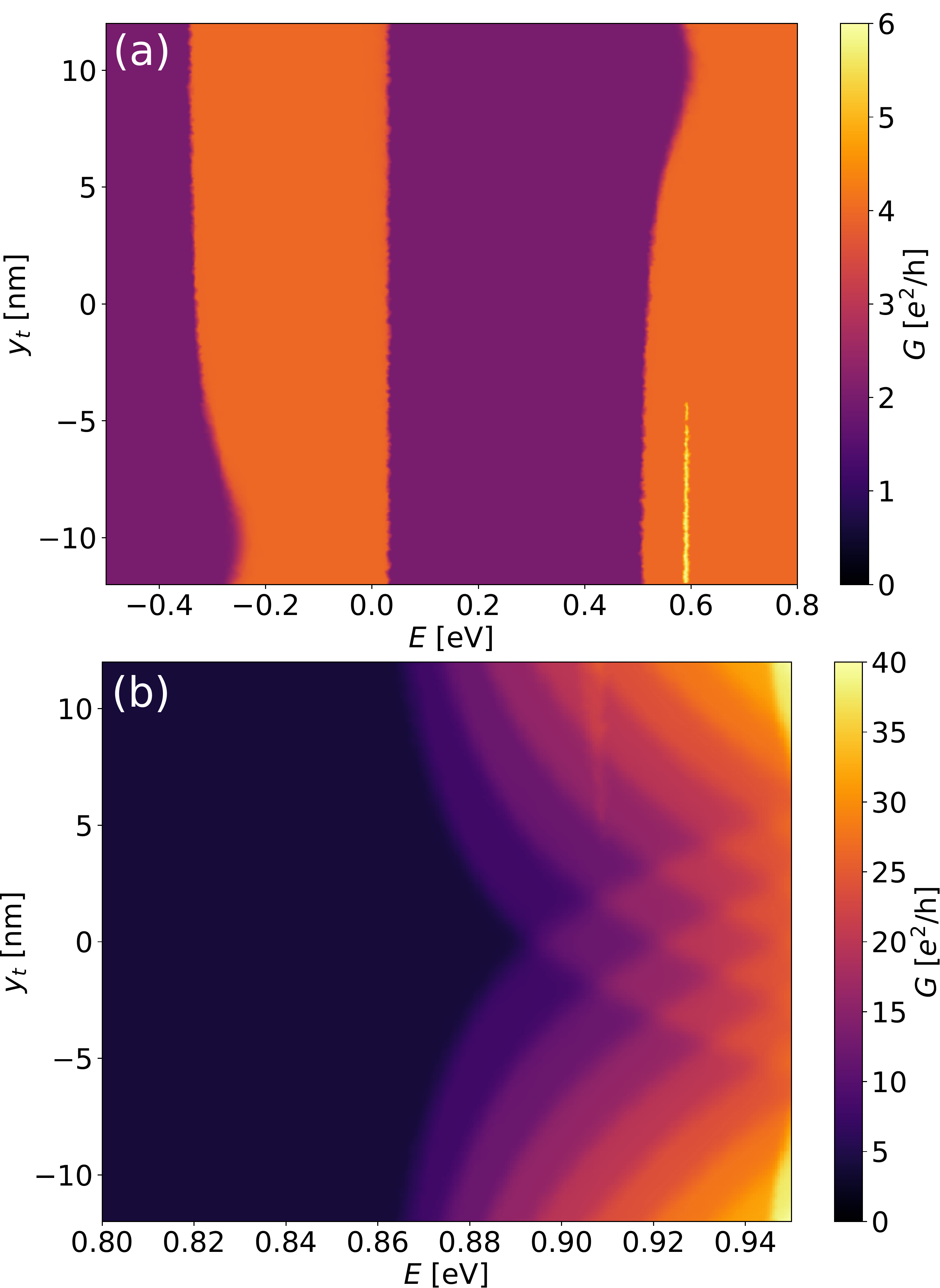}
\caption{Conductance map versus the tip position in the $y$-direction (for $x_t$ = 0) and the Fermi energy for a zigzag nanoribbon for edge (a) and (b) bulk dominated transport.}
\label{pristine_map}
\end{figure}

Let us first focus on the Fermi energy range where the transport is dominated by the edge modes. In Figure \ref{pristine_map} (a) we show conductance of the ribbon versus the tip position along the $y$ axis ($y_t$) and the Fermi energy in range $[-0.5,0.8]$ eV. The considered ribbon is invariant in the $x$ direction and we set $x_t$ to zero. 

\begin{figure}[ht!]
\center
\includegraphics[width = 13cm]{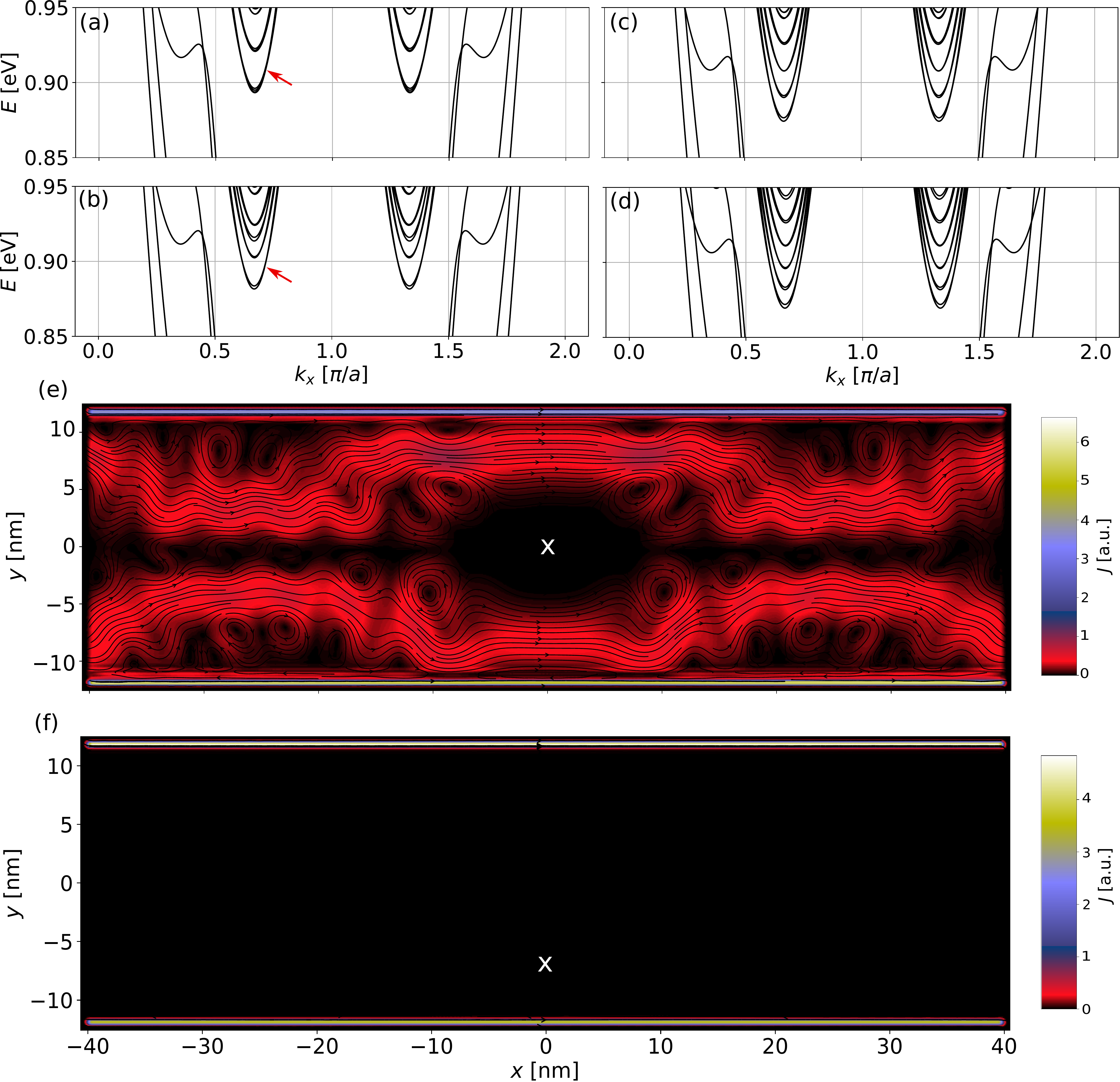}
\caption{(a)-(d) Dispersion relations close to the conduction band minimum for four positions of the tip with $y_t = 0$ (a), $y_t = 3$ nm (b), $y_t = 6$ nm (c), $y_t = 9$ nm (d). (e) Density current probability obtained for the tip positioned at the center of the ribbon (marked with the white cross) for $E=0.9$ eV. The blue and yellow colors correspond to the current density localized at the Mo and S edges respectively. (f) Density current probability obtained for the tip positioned at $y_t = -6$ nm (marked with the white cross) for the Fermi energy $E=-0.3$ eV where the transport occurs solely through the edges. The current flows from left to right and its density takes zero values in the bulk of the ribbon (black) and it is maximal on the edges (yellow). In the panels (e) and (f) the black arrows denote the current direction.}
\label{RD_current}
\end{figure}

In the absence of the tip the conductance is proportional to the number of current carrying modes. The lack of the energy gap in the metallic band structure of a pristine zigzag ribbon leads to nonzero conductance in the whole map. For the energy range considered in Figure \ref{pristine_map}(a) this corresponds to the current being transmitted at the edges of the ribbon -- see Figure \ref{RD_current}(f). We distinguish two main regimes in the map (depicted with violet and orange colour) where the conductance takes values $2e^2/h$ and $4e^2/h$ due to Kramers degeneracy. The violet colour in the map  corresponds to the case when a single (Mo or S) edge is populated, while the orange colour denotes the transport through both the edges \cite{ridolfi_electronic_2017}. A prominent feature of the map is the stability of the conductance when the tip approaches the edges despite the fact that the current is transmitted through them. Only when the Fermi energy is tuned near the bottom of the edge bands we observe that the conductance drops by $2e^2/h$ as it happens for $E = -0.3$ eV and $E = 0.55$ eV. For such values of $E$ the small potential perturbation induced by the tip locally lifts up the energy of the edge modes. As a result their band bottom positions above the Fermi energy which in turn blocks the transport through the S edge for $E=-0.3$ eV and Mo terminated edge for $E = 0.55$ eV. As a result the incoming electron is reflected completely from the depletion region generated by the tip which correspondingly lowers the conductance by the conductance quanta.

In the map we also observe a sharp peak of the increased conductance due to the bend Mo edge band at energies close to  $0.6$ eV [see the dispersion relation in Figure \ref{bandstructure}] which results in propagation of two edge modes on this side of the ribbon.

In Figure \ref{pristine_map}(b) we present the conductance map versus the Fermi energy and the tip position along the $y$ axis for the energies tuned to the conduction band. Here the current is carried mainly by the modes that have either maximum in the center of the ribbon (states with even parity) or maxima symmetrically around the center (odd states). Nevertheless, we see that actually the conductance mapping does not probe the probability current distribution, but rather we observe a set of fans of quantized conductance which increases when the tip is moved towards the edges of the ribbon---see e.g. the conductance at $E=0.88$ eV. 

The above observation can be explained by an analysis of the energy bands of the ribbon section in which the tip is located. Here, the energy values are dominated by the energy separation between states of transverse quantization in the ribbon which bands we find in the top part of Figure \ref{bandstructure}. Taking a cross-section of the system at $x=0$ and assuming its invariance along the $x$ direction we calculate the band structure for different $y_t$'s, as we moved the tip toward the top edge, and plot it in Figures \ref{RD_current}(a)-(d). When the tip is positioned in the center of the ribbon with $y_t=0$ we obtain the dispersion relation presented in Figure \ref{RD_current}(a). We observe two sets of parabolas that correspond to $K$ and $K'$ modes. Each band in those sets is nearly fourfold degenerate -- see the red arrow in Figure \ref{RD_current}(a). The degeneracy results from the presence of two bands split by SOC and two spatial channels for the current flow---on the two opposite sides of the tip. We show the probability current at $E = 0.9$ eV in Figure \ref{RD_current}(e), where the colors corresponds to the magnitude of the current probability. In the plot we observe the current flowing around the tip. Note that the degeneracy of the current carrying modes at the opposite sides of the tip is not perfect due to coupling to two non-equivalent edges terminated by Mo and S atoms [see the edge current flow in Figures \ref{RD_current}(e)(f)]. When the tip is moved towards the edge of the ribbon, the band structure gradually changes as can be seen in Figures \ref{RD_current}(a)-(d). Due to widening of the propagation channel on one side of the tip and narrowing the channel on the other side the energies of the modes propagating in the wider (narrow) channel decrease (increase). Correspondingly, the fourfold degeneracy of the bands is replaced by twofold degeneracy due presence of two spin modes -- see the red arrow in Figure \ref{RD_current}(b). As a result, subsequent bands cross the Fermi energy when the tip is moved towards the edge and accordingly the conductance gradually increases by $2e^2/h$ steps.

\subsection{Coherent electron flow from a QPC}
\begin{figure}[ht!]
\center
\includegraphics[width = 12cm]{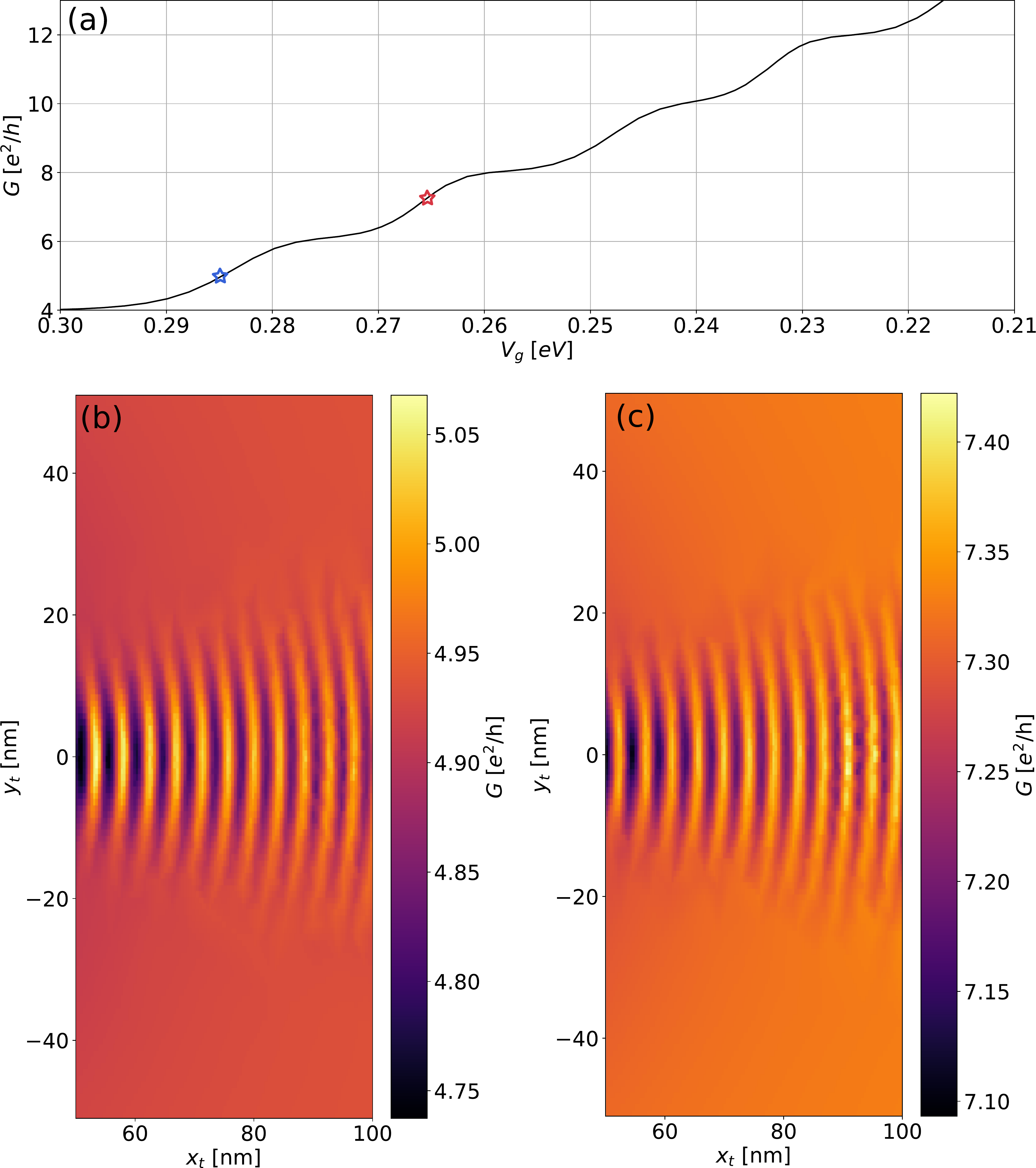}
\caption{(a) The conductance versus QPC potential. The SGM conductance map on the first (b) and (c) the second step, as indicated by a blue and a red star in panel (a), respectively. The results obtained within the tight-binding model for $E = 0.9$ eV.}
\label{TB_branching}
\end{figure}

The coherent electron scattering can be most distinctly demonstrated by a SGM conductance mapping of the electron flow from a QPC. Let us focus now on the conductance mapping in a $\mathrm{MoS_2}$ ribbon with a QPC. We consider a model of a QPC in the form of a split-gate as studied in the experiment \cite{sharma_split-gated_2017} with the potential plotted in Figure \ref{system} on the colourmap. We consider a wide zigzag ribbon with the width of 200.2 nm and present in Figure \ref{TB_branching}(a) the quantized conductance versus the QPC potential for $E=0.9$ eV. For the most positive values of the gate potential in Figure \ref{TB_branching}(a) we see that the conductance reaches the value of $4e^2/h$. We have checked that this plateau is very stable in $V_g$ and results from the edge modes that are pinched off only if the gates create potential barrier high enough to lift up the edge mode band above the Fermi energy.

\subsubsection{Branching versus SOC strength.}
Let us now focus on two $V_g$ values for which we denote the conductance by stars in Figure \ref{TB_branching}(a). In the bottom panels we plot corresponding SGM conductance maps versus the position of the scanning probe. We observe that in the both cases the current flows in the form of a single branch. This in the striking contrast with the measurements of 2DEG in semiconductor heterostructures \cite{topinka_coherent_2001} where in absence of magnetic field, the number of branches $N$ reflects the conductance value quantized in $N\times2e^2/h$. This is the hallmark of transmission through the QPC constriction of modes with increased quantization number and by that increased number of maxima in the cross-section of the probability current in the transverse direction.

\begin{figure}[ht!]
\center
\includegraphics[width = 10cm]{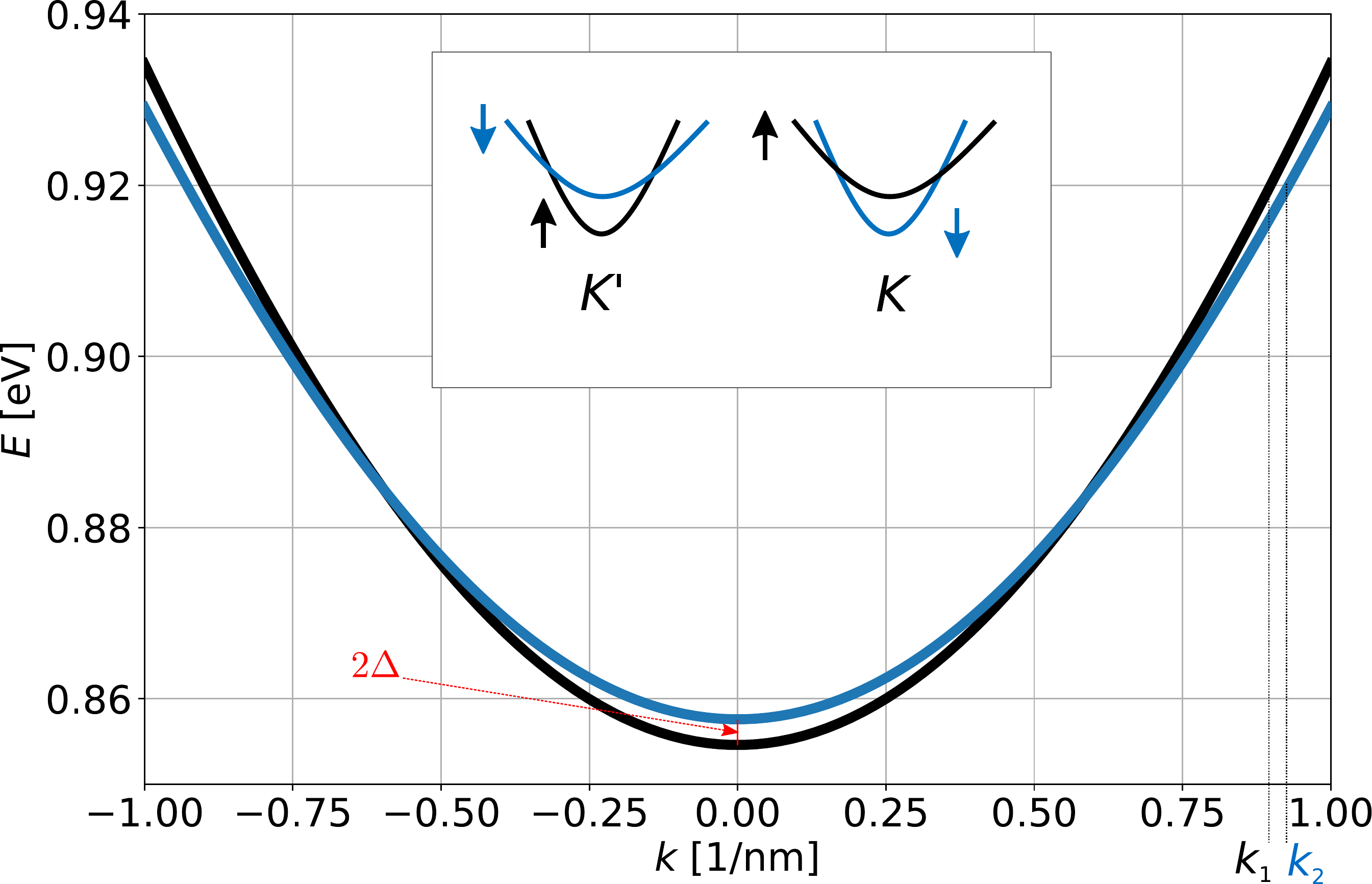}
\caption{The band structure of the bulk $\mathrm{MoS_2}$ monolayer for $k_y=0$ obtained in the continuum model with two possible values of the wavevector for the modes with positive group velocity at $E=0.92$ eV denoted with $k_1$ and $k_2$. The inset shows schematically the spin- and valley-split bands in the conduction band of the monolayer.}
\label{cnt_bandstructure}
\end{figure}

It is important to realize here that in a bulk TMDC monolayer the conduction band minimum consists of two spin-opposite modes localized at two sets of nonequivalent $K$ and $K'$ points in the Brillouin zone. At each $K$ and $K'$ points there are two bands that have minima separated by the energy of $2\Delta$ stemming from the intrinsic SOC---see the inset to Figure \ref{cnt_bandstructure}. As a result, in a nanoribbon the resulting dispersion relation consist of two sets of nearly-parabolic bands for $K$ and $K'$ polarized modes where in each set there are bands corresponding to spin-up and spin-down modes separated by energy of $2\Delta$. The commonly assumed value of the spin-orbit gap of a freestanding $\mathrm{MoS_2}$ sheet is $2\Delta=3$ meV as obtained from DFT calculations \cite{kormanyos_spin-orbit_2014}. However, recent experiments suggest that in fact $\Delta$ can be different, and sample dependent: in the experimental measurements of Shubnikov-de Haas oscillations in $\mathrm{MoS_2}$ the gap value was found to be 15 meV \cite{pisoni_interactions_2018}. Also, for $\mathrm{MoSe_2}$  the experimentally probed spin-orbit splitting turned out to be larger than predicted theoretically \cite{larentis_large_2018}. Those measurements suggest that gating and the presence of substrate  might affect the magnitude of spin-orbit splitting in the conduction band. In the QPC constriction the transverse mode quantization energy of order of meV is comparable to $2\Delta = 3$ meV. In the following we inspect how the strength of SOC affects the SGM conductance mapping by varying the value of $\Delta$.

To investigate the physics of the observed branched electron flow in a more detail let us approximate the monolayer band structure close to the conduction band edge at the $K$ points by the continuum Hamiltonian \cite{kormanyos_spin-orbit_2014}, 

\begin{equation}
H_{\mathrm{continuum}}=\left(
\begin{array}{cccc}
\frac{\hbar^2\mathbf{k}^2}{2m^{+\uparrow}} + \Delta & \Gamma & 0 & 0 \\
\Gamma & \frac{\hbar^2\mathbf{k}^2}{2m^{-\uparrow}} - \Delta & 0 & 0 \\
0 & 0 & \frac{\hbar^2\mathbf{k}^2}{2m^{+\downarrow}} - \Delta & \Gamma \\
0 & 0 & \Gamma & \frac{\hbar^2\mathbf{k}^2}{2m^{-\downarrow}} + \Delta \\
\end{array}\right) -\mu\mathbf{1}.
\end{equation}
We take the effective masses: $m^{-\uparrow} (m^{+\downarrow})= 0.44 m$ (where $m$ is the free electron mass) that corresponds to $K'$ ($K$) spin-up (spin-down) bands and $m^{-\downarrow} (m^{+\uparrow})= 0.49 m$ that corresponds to $K'$ ($K$) spin-down (spin-up) bands. We neglect trigonal warping \cite{kormanyos_monolayer_2013} and set the chemical potential $\mu=-0.8561$ eV to match the band structure obtained in the tight-binding model. $2\Delta$ is the band splitting due to the intrinsic SOC and $\Gamma$ is the valley mixing parameter. The main advantage of this model is the ability to explicitly control the strength of internal the SO coupling and the valley mixing. The resulting band structure for $\Delta = 1.5$ meV is presented in Figure \ref{cnt_bandstructure}.

\begin{figure}[ht!]
\center
\includegraphics[width = 12cm]{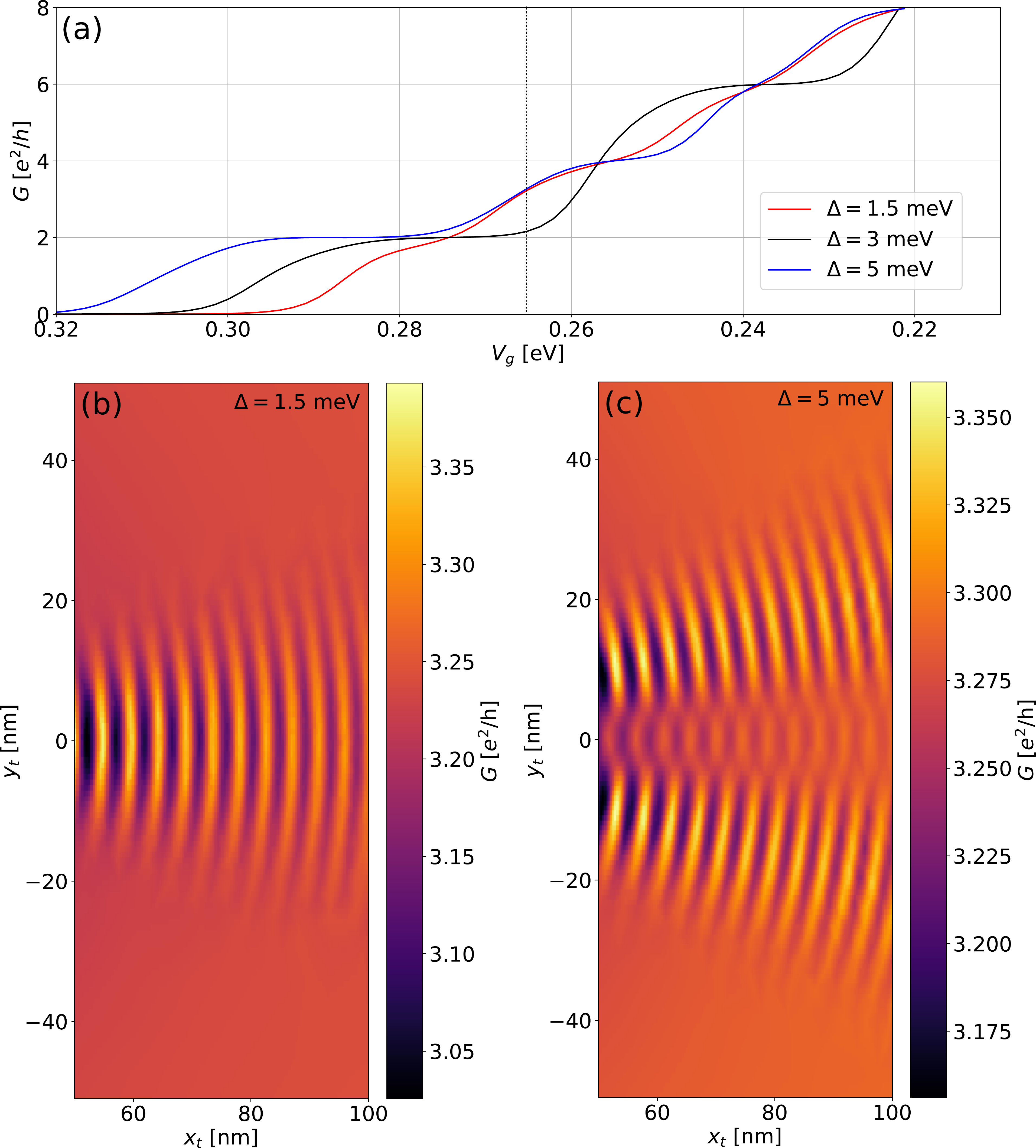}
\caption{(a) Conductance versus the QPC potential for three values of the internal SOC strength $\Delta$. (b) and (c) present the SGM mapping on the second step. The results are obtained in the continuum model approximating the tight-binding description near the conduction band minimum -- see text for details -- for $E=0.9$ eV.}
\label{cont_G_Vg_maps}
\end{figure}

We use the continuum model to describe the same system as previously treated in the tight-binding approach, i.e. we perform transport calculations describing a nanoribbon with the QPC and SGM potentials using the Hamiltonian $H = H_{\mathrm{continuum}} + V\bm{1}$, where $\bm{1}$ is an identity matrix. We discretize the Hamiltonian $H$ on a square mesh with the lattice constant $\Delta x = \Delta y=1$ nm and apply a semi-open boundary condition analogously as in the tight-binding model case. 

The obtained conductance traces versus the QPC potential are plotted in Figure \ref{cont_G_Vg_maps}(a) for three values of the SOC strength $\Delta$. For $\Delta = 1.5$ meV we find that the conductance curve depicted in panel (a) reproduces the trace obtained in tight-binging model [Figure \ref{TB_branching}(a)] differing only by the offset $4e^2/h$ as the edge modes are not captured in the continuum approximation. Accordingly, when the QPC is set to the second step, there is a single branch in the conductance map of Figure \ref{cont_G_Vg_maps}(b) similar to the one presented in Figure \ref{TB_branching}(c).

In Figure \ref{cont_G_Vg_maps}(a) with the black and blue curve we plot the conductance for $\Delta=3$ meV and $\Delta = 5$ meV. We observe that for $\Delta=3$ meV  when $V_g$ decreases the conductance raises from $2e^2/h$ already to $6e^2/h$ which reflects the fact that the QPC transmits three spin-degenerate modes. When we further increase $\Delta$ to $5$ meV the overall shape of the conductance trace become again similar to the one obtained for $\Delta = 1.5$ meV. However, now mapping the current flow on the second step we observe that the current divides in two branches [Figure \ref{cont_G_Vg_maps}(c)]. Comparing the results for $\Delta=1.5$ meV and $\Delta = 5$ meV it becomes clear that in the former case the first plateu corresponds to the transmission of two ground states of the transverse quantization with the $K'\uparrow$, $K\downarrow$ polarization and the value of $4e^2/h$ is obtained as the QPC transmits four spin- and valley- opposite modes ($K\uparrow$, $K'\downarrow$, $K\downarrow$, $K'\uparrow$) also in the ground state of the transverse excitation. For a strong spin splitting the spin-orbit split bands ($K'\downarrow$, $K\uparrow$) are shifted to higher energies and now the conductance of $4e^2/h$ on the second step corresponds rather to the transmission of a ground state and an excited state of transverse quantization of two spin- and valley- opposite modes ($K'\uparrow$, $K\downarrow$). Therefore, the SGM mapping of the branched electron flow provides a tool to distinguish the nature of the conductance quantization by the QPC.  

\subsubsection{Oscillation period in the presence of spin- and valley-split bands.}
\begin{figure}[ht!]
\center
\includegraphics[width = 12cm]{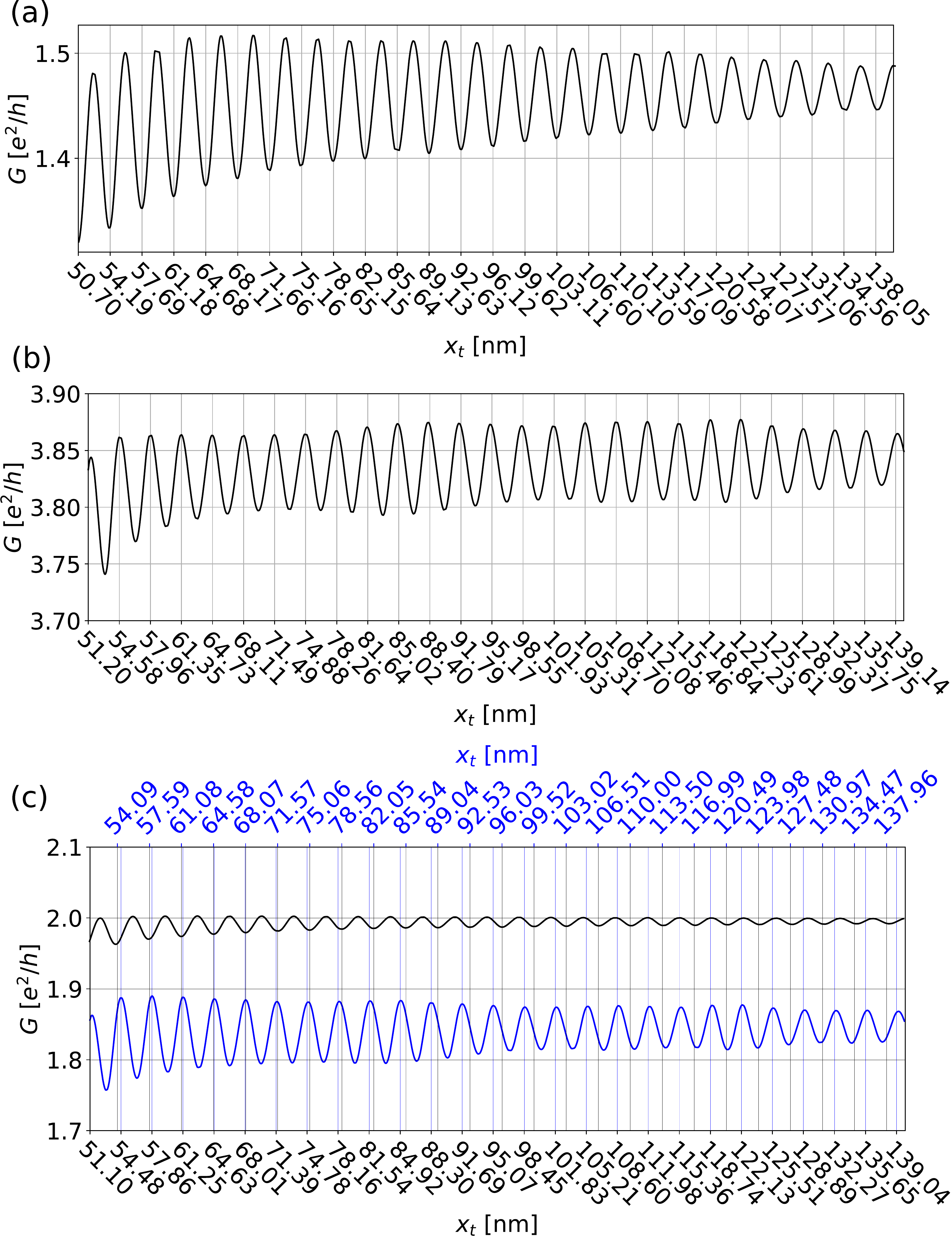}
\caption{(a) Conductance oscillations with the period of $3.494$ nm corresponding to the mode with wavevector $k_1$ obtained at the first QPC conductance step. (b) Conductance oscillations at the second QPC conductance step. (c) The blue (black) curve depict the conductance from $K\downarrow$, $K'\uparrow$ modes ($K\uparrow$, $K'\downarrow$). The two curves oscillate with period $\pi/k_1$ and $\pi/k_2$. The sum of this oscillations gives the trace presented in the panel (b). The results are obtained for $E = 0.92$ eV.}
\label{oscillations}
\end{figure}

An important feature of the conductance maps presented in Figure \ref{TB_branching} are the radial fringes--conductance oscillations obtained when the tip moves outwards from the QPC narrowing. For a single band semiconductor they are separated by a half of Fermi wavelength and result from the interference of the wave function of the electron exiting the constriction and the wave reflected back to it \cite{leroy_imaging_2005}.

To inspect the period of oscillations we consider now a modified version of the system described in the continuum model, i.e. we consider a wide (500 nm) ribbon for which we set to zero the potential created by the QPC gates for $x>50$ nm. This way we exclude all possible sources of back scattering for waves exiting the QPC and make sure that the QPC does not affect the local chemical potential outside the region close to the QPC gates. We consider the Fermi energy $E = 0.92$ eV and $\Delta = 1.5$ meV.

In Figure \ref{cnt_bandstructure} we plot the band structure of a bulk $\mathrm{MoS_2}$ flake near the conduction band minimum obtained in the continuum model. At the Fermi energy of $0.92$ eV there are two possible values of the electron wavevector denoted with $k_1$ and $k_2$ that correspond to spin- and valley opposite modes. When the QPC is tuned to the first conductance plateau it transmits the band with minimal energy at $k_x = 0$, i.e. the one that gives the wavevector $k_1$ at energy 0.92 eV. Accordingly, when we inspect the conductance oscillations presented in Figure \ref{oscillations}(a) we find that they have a period corresponding to $l_1=\pi/k_1$. On the other hand, if we monitor the oscillations at the second conductance step plotted in Figure \ref{oscillations}(b), we observe that there is no single oscillation period. In Figure \ref{oscillations}(c) we plot conductance oscillations at the second step divided into components corresponding to the ground state in the band structure (with the wavevector $k_1$ at the energy 0.92 eV) [blue curve] and to the first excited state (with the wavevector $k_2$ at the energy 0.92 eV) [black curve]. They have two different oscillation periods and their composition gives the  slightly beating curve of Figure \ref{oscillations}(b). This is the hallmark of valley and spin coupling specific to TMDCs.

\subsubsection{Beating pattern due to valley mixing.}
It is important to note here that despite the electrons propagate in a composition of the two valley states there is no inter-valley scattering -- the electrons are injected and exit the system in a well defined valley state. Correspondingly, each valley state exiting the QPC is interfering with its counterpart reflected from the tip. It is known that short range scatterers as vacancies in the atomic lattice \cite{wakabayashi_perfectly_2007, wakabayashi_electronic_2009, lima_effects_2012} might lead to the inter-valley scattering provided the Fermi energy is high enough to allow for spin-preserving valley flips \cite{gut_valley_2019} and hence to mixing of the valley states propagating through the ribbon. Alternatively, the valley scattering with spin-flips can be caused by magnetic impurities \cite{avalos-ovando_symmetries_2016}.

\begin{figure}[ht!]
\center
\includegraphics[width = 10cm]{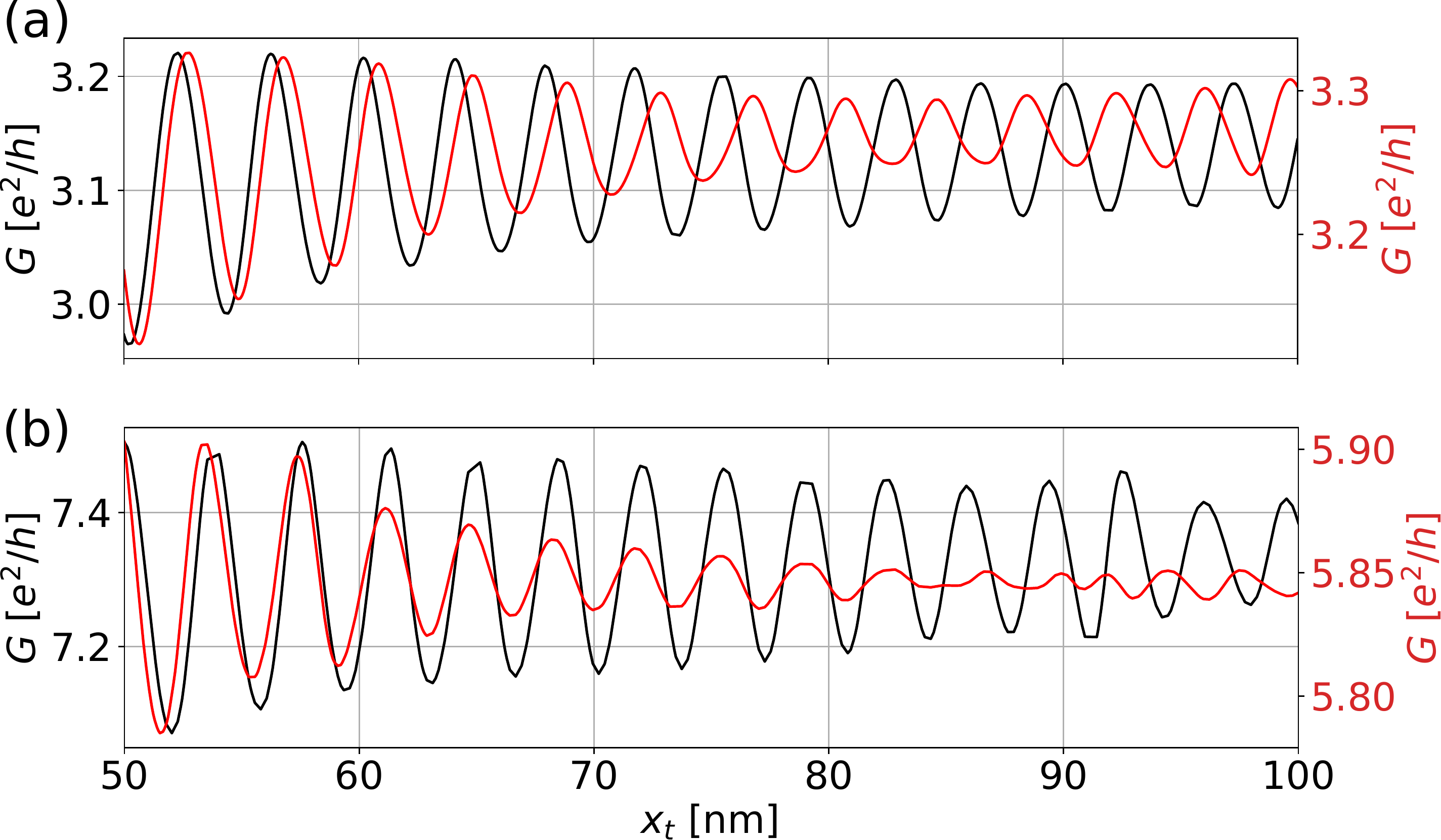}
\caption{(a) Conductance oscillations as a function of the longitudinal tip position  obtained in the continuum model on the second QPC conductance step without (black) and with (red) valley mixing. (b) Conductance oscillations as a function of the longitudinal tip position obtained in the tight-binding model on the second QPC conductance step without (black) and with (red) atomic impurities. The results are obtained for $E=0.92$ eV.}
\label{beating}
\end{figure}

Let us now investigate the effect of the valley mixing on the conductance oscillations. We first analyze the results obtained in the continuum model. In Figure \ref{beating}(a) we show the conductance calculated in the continuum model with $\Gamma=0$ and $\Gamma=1.5$ meV with black and red curves respectively. We observe that the valley mixing induced by the nonzero $\Gamma$ parameter results in a strong beating pattern in the conductance.

Finally, we go back to the system described in the tight binding model. To test that we introduce 300 vacancies in the region span by $x = [-25,50]$ nm and $y=[-50,50] $ nm \cite{note}. The resulting conductance trace is shown in Figure \ref{beating}(b) with the red curve. Comparing it to the one obtained in the system with no disorder (black) we note a significant beating pattern due to valley mixing made possible now by the disorder.

\section{Summary and conclusions}
In summary, we have studied the electronic transport in a $\mathrm{MoS_2}$ ribbon in the presence of a scanning probe. We adopted the the tight-binding and the continuum approaches for the description of TMDCs monolayer nanodevices. For a pristine ribbon we showed that the edge modes can be mapped by the SGM technique provided the Fermi energy is tuned to the bottom of their bands. When the ribbon is doped into conduction band the conductance probing reveals fan pattern in the conductance maps as the tip scans across the sample. This is due to the presence of the quantized spin- and valley- coupled modes in the ribbon. In wide structures where the conductance is controlled by a split-gate QPC we demonstrated that the current exits the constriction in branches which number is controlled not only by the QPC constriction itself as in ordinary 2DEGs but also by the intrinsic SOC strength. We explained that the conductance oscillation fringes evidence of two possible wavevectors for the charge carriers due to the SOC splitting of the bands. Finally, we showed that valley mixing induced by the short-range scatterers induces significant beating in the conductance oscillations.

\ack
We acknowledge support within POIR.04.04.00-00-3FD8/17 project carried out within the HOMING programme of the Foundation for Polish Science co-financed by the European Union under the European Regional Development Fund. The calculations were performed on PL-Grid Infrastructure. The authors acknowledge helpful discussions with T. Rosdahl, A. Akhmerov and M. Wimmer.

\section*{References}
\bibliographystyle{iopart-num}
\bibliography{MoS2_SGM}

\begin{thebibliography}{10}
\bibitem{note}
 {\em The edge vacancies can significantly alter the transport properties of a
  ribbon when the Fermi energy is tuned into the bulk gap as found in
  \cite{ridolfi_electronic_2017}. There, the authors found that even a single
  edge vacancy can significantly alter the edge transport, provided that the
  sites close to the edges are vacant. Nevertheless, in our SGM study of the
  system with QPC, the conductance mapping is sensitive to the current flow
  from the QPC constriction -- far from the edges. The disturbance of the edge
  transport modifies the conductance---when the edge transport is blocked---but
  that alters the overall conductance magnitude, not its change induced by the
  tip which scans the vicinity of the QPC constriction\/}
\end{thebibliography}
\end{document}